\documentstyle[11pt]{article}
\begin{document}
%%%%%%%%%%%%%%%%%%%%%%%%%%%%%%%%%%%%%%%%%%%%%%%%%%%%%%%%
%     GENERIC USEFUL DEFINITIONS                       %
%%%%%%%%%%%%%%%%%%%%%%%%%%%%%%%%%%%%%%%%%%%%%%%%%%%%%%%%
\newcommand{\DUR}{\em Department of Physics\\
                      University of Durham \\
                      Science Laboratories, South Road\\
                      Durham, DH1 3LE, U.K.}

\newcommand{\NPB}[3]{{ Nucl.\ Phys.\/} {\bf B#1} (19{#3}) {#2}}
\newcommand{\PLB}[3]{{ Phys.\ Lett.\/} {\bf #1B} (19{#3}) {#2} }
\newcommand{\PRD}[3]{{ Phys.\ Rev.\/}  {\bf D#1} (19{#3}) {#2} }
\newcommand{\PRA}[3]{{ Phys.\ Rev.\/}  {\bf A#1} (19{#3}) {#2} }
\newcommand{\PRL}[3]{{ Phys.\ Rev.\ Lett.\/} {\bf #1} (19{#3}) {#2} }
\newcommand{\ZFP}[3]{{ Zeit.\ f.\ Phys.\/}, {\bf C #1} (19{#3}) {#2} }
\newcommand{\IJA}[3]{{ Int.\ J.\ Mod.\ Phys.\/} {\bf A#1} (19{#3}) {#2} }

\def\lhs{{\it l.h.s.\/ }}
\def\rhs{{\it r.h.s.\/ }}
%%%%%%%%%%%%%%%%%%%%%%%%%%%%%%%%%%%%%%%%%%%%%%%%%%%%%%%%%%
%   LATEX-STRUCTURE DEFINITIONS				 %
%%%%%%%%%%%%%%%%%%%%%%%%%%%%%%%%%%%%%%%%%%%%%%%%%%%%%%%%%%
\def\bqn{\begin{equation}}
\def\eqn{\end{equation}}
\def\bqna{\begin{eqnarray}}
\def\eqna{\end{eqnarray}}
\def\nn{\nonumber}
\def\bit{\bibitem}
\def\vs{\vspace{.25in}}
\def\thefootnote{\fnsymbol{footnote}}

\def\QQa{\renewcommand{\baselinestretch}{1.3}\Huge\large\normalsize}

%%%%%%%%%%%%%%%%%%%%%%%%%%%%%%%%%%%%%%%%%%%%%%%%%%%%%%%%%%%
%       DEFINITION OF SECTEQNO                            %
%%%%%%%%%%%%%%%%%%%%%%%%%%%%%%%%%%%%%%%%%%%%%%%%%%%%%%%%%%%
\makeatletter
\def\secteqno{\@addtoreset{equation}{section}%
\def\theequation{\thesection.\arabic{equation}}}

%To disable this
\def\endsecteqno{\def{theequation\{\@ifundefined{chapter}%
{\arabic{equation}}{\thechapter.\arabic{equation}}}}
\makeatother

%%%%%%%%%%%%%%%%%%%%%%%%%%%%%%%%%%%%%%%%%%%%%%%%%%%%%%%%%%%
%       DEFINITIONS USEFUL FOR THIS PAPER                 %
%%%%%%%%%%%%%%%%%%%%%%%%%%%%%%%%%%%%%%%%%%%%%%%%%%%%%%%%%%%
\def\gl{\tilde{g}}
\def\mg{m_{\gl}}
\def\as{\alpha_s}
\def\asmz{\alpha_s(M_Z)}
\def\GeV{{\rm GeV}}
\def\ycut{y_{\rm cut}}
\def\thetanr{\theta^*_{\mbox{\sc nr}}}

%%%%%%%%%%%%%%%%%%%%%%%%%%%%%%%%%%%%%%%%%%%%%%%%%%%%%%%%%%%
%       THE FRONT PAGE                                    %
%%%%%%%%%%%%%%%%%%%%%%%%%%%%%%%%%%%%%%%%%%%%%%%%%%%%%%%%%%%
\pagestyle{empty}
{\hfill \parbox{6cm}{\begin{center} DTP/93/72 \\
                                    September 1993
                      \end{center}}}

\vspace*{2cm}
\begin{center}
\large{\bf LIGHT GLUINOS IN FOUR-JET EVENTS AT LEP} \\
\vskip .6truein
\centerline {R. Mu\~noz-Tapia \footnote{e-mail: RMT@HEP.DUR.AC.UK}
              and W.J. Stirling \footnote{e-mail: WJS@HEP.DUR.AC.UK}}

\end{center}
\vspace{.3cm}
\begin{center}
\DUR
\end{center}
\vspace{1.5cm}

\centerline{\bf Abstract}
\medskip
 The light gluino hypothesis can explain the apparent incompatibility between
 the measurements of $\as$ at low- and high-energy. Such gluinos are
produced directly
 in four-jet events, for which we perform a detailed analysis. Because
the jet energies are not large, the  effect of the non-zero gluino mass
is important. We take the
gluino mass into account in the computation of the cross sections and shape
 variables. As expected, we find that mass effects tend to reduce the
impact of the gluinos in the cross section, weakening the bounds from
obtaining assuming massless gluinos.
\newpage

%%%%%%%%%%%%%%%%%%%%%%%%%%%%%%%%%%%%%%%%%%%%%%%%%%%%%%%%%%%%%
%          THE BODY  OF THE PAPER                           %
%%%%%%%%%%%%%%%%%%%%%%%%%%%%%%%%%%%%%%%%%%%%%%%%%%%%%%%%%%%%%
\pagestyle{plain}
\QQa
In recent  years, many precision tests of QCD have been carried out at
LEP and, in particular,
 much attention has been devoted to the measurement of the
strong coupling constant at the $M_Z$ scale, $\asmz$. These
measurements  can be compared with the  values extracted from
deep inelastic experiments, at lower energies, and evolved using the
standard QCD renormalization group equation. It has already been noted
that there is a slight discrepancy between the results obtained in this
way. The LEP measurements  yield an average of $\asmz = 0.122 \pm
0.006$, while the deep inelastic values suggest $\asmz = 0.112 \pm
0.005$ \cite{alpha}. Of course the statistical significance of this
discrepancy is very small, but nonetheless it has led to speculation
that the evolution of coupling is being slowed down by a contribution
to the $\beta$-function of a new light, neutral, coloured fermion, the
so-called
\lq light gluino' hypothesis \cite{kuhn}. Although it is theoretically
difficult to reconcile such an object with a realistic supersymmetric
Standard Model,  there is in fact an experimental window open precisely
in the few GeV region \cite{gluino-window}. A gluino of this mass, in the
adjoint
representation, would slow the evolution between the deep inelastic and
LEP scales by just the correct amount  to reconcile the $\asmz$
mesurements.  In simple terms,
\begin{eqnarray}
{{\rm d} \alpha_s(\mu) \over {\rm d} \log\mu} & = &
  \beta_0 \alpha_s^2(\mu) \nonumber \\
\beta_0   &  =  & {1\over 2 \pi} \left[ 11 - {2\over 3} n_f  - 2 \theta(\mu -
\mg)  \right]  \; .
\end{eqnarray}
Note that the above threshold, the effect is the same  as
increasing the number of quark flavours, $n_f \to n_f + 3$.
Of course the presence  of a light gluino also modifies the values
extracted for $\alpha$, but the effects are fairly  small compared to the
experimental uncertainties on the measured $\alpha_s$ values \cite{ellis2}.

%\begin{figure}
%\centerline{
%\inclps{fig1.ps}{500 pt}{650 pt}{50 pt}{300 pt}{600} }
%\caption{Feynman diagrams contributing to the four-jet cross section.
%         Other permutations are not shown.}
%\end{figure}
It is certainly worth looking for evidence  of the light gluino in other
processes.
At LEP,  gluino pairs can be produced directly at $O(\alpha_s^2)$, {\it i.e.}
as a contribution to the  four-jet cross section \cite{farrar}. One can,
at least in principle, extract from the data the number of light
hadronic fermion
pairs contributing to $e^+e^- \to q \bar q f \bar f$. Naively,  a light
gluino would increase $n_f$ by three, just as for the evolution of the
coupling constant.
The main purpose of this note is to analyse the four-jet
event rate at LEP energies, in order to quantify as the effect due to
gluino production. In particular, we are primarily interested in the effect of
the
non-zero mass $\mg$ on the cross section. Previously analyses
\cite{farrar,aleph,opal,hebbeker} have assumed massless quarks, gluons and
gluinos, but since the
energies of the gluino jets are not large, mass effects will presumably be
important.

In Fig.\ 1(a)-(d) we show the lowest order Feynman graphs for 4-jet production
in QCD. The  gluino  contributes through gluon splitting
processes of type (d), shown in  Fig.\ 2.
To study the effect of a non-zero gluino mass,  we have computed the
matrix element  for $e^+e^- \to q \bar q \gl \bar{\gl}$ with $\mg \neq
0$, using the spinor techniques of reference \cite{spinor}.
%\begin{figure}
%\centerline{
%\inclps{fig2.ps}{390 pt}{660 pt}{200 pt}{550 pt}{600} }
%\caption{Feynman diagram corresponding to the the production of a pair of
%          gluinos in four-jet events. The other permutation is not shown.}
%\end{figure}
In Fig.\ 3
we show (solid line) the total cross section $\sigma(Z\to \sum_q q \bar q
\gl\bar{\gl})$, normalized to the leading order (two-jet) cross section
$\sigma_0 \equiv \sigma(Z\to \sum_q q\bar q)$, as a function of $\mg$.
Note that this cross section is infra-red finite for $\mg > 0$.
For $\mg > O(5\ \GeV)$, the cross section falls exponentially with the
gluino mass. In order to define the part of this which corresponds to
the four-jet cross section we need to introduce a jet algorithm.
In what follows we will adopt the widely-used JADE algorithm, {\it i.e.}
we introduce a dimensionless parameter $\ycut$ and require that jets
$i$ and $j$ be separated in phase space according to
\bqna
\tilde{y}_{ij}&>&\ycut \nn \\
\tilde{y}_{ij}&=&\frac{2 E_i E_j (1-\cos\theta_{ij})}{s},
\eqna
where $\theta_{ij}$ is the angle between the jets with energies $E_i$ and
$E_j$ respectively. In our calculation, the
indices $ 1 \leq i,j \leq 4$ run over  the four final-state
partons. Fig.\ 3 shows the gluino pair contribution to the four-jet
cross section defined in this way, for three $\ycut$ values, again as a
function of $\mg$.  Notice that for $\ycut=0.01$,
the cross section decreases by a factor of two going from $\mg=0$ to
$\mg=5\ \GeV$.
%\begin{figure}
%\centerline{
%\inclps{fig3.ps}{505 pt}{643 pt}{60 pt}{160 pt}{700}  }
%\caption{Mass dependence of the 4-jet total cross section for
%          different $\ycut$. The values are normalized to the
%          lowest order cross section.}
%\end{figure}

In Fig.~4 the various contributions to the total four-jet cross section
({\it i.e.} summed over the processes shown in Figs.~1 and 2)  are
presented, as a function of $\ycut$.
As can be seen, by far the most important contribution is from
the $q \bar q  g g $ final state. The four-quark contribution $q \bar q
 q \bar q$ is
one order of magnitude smaller. Note that in computing this
contribution,  the $b$ quark mass has
been taken into account.
The gluino contribution is shown for
different masses. For $\mg =0$ it is of the same order as the quark
contribution ---  the
number of flavours is compensated by the enhanced colour factor of the gluino.
A non-zero  mass has, however, an important effect in suppressing
 the cross section:
the value for $\mg=10$ GeV is four times lower than that for $\mg=5$ GeV at
$\ycut=0.01$. The conclusion from this Fig.~4 is that a heavier gluino
would be very hard to detect,
even if one could separate the  fermion from the vector boson jets.
%\begin{figure}
%\centerline{
%\inclps{fig4.ps}{525 pt}{643 pt}{ 40 pt}{160 pt}{700} }
%\caption{$\ycut$ dependence of the total cross section for
%        the gluon, quark, and gluino contributions to the 4-jet final
%        state.
%        In the quark line, the
%         mass effect of the $b$-quark is taken into account.}
%\end{figure}

Reference~\cite{aleph} describes an attempt by the ALEPH collaboration
 to measure the QCD
colour factors from a sample of 4-jet events.
The idea is to fit the theoretical predictions to
the data,  leaving the colour factors to be determined by the fit.
The theoretical expression for the  $Z\to q\bar{q}gg$ contribution,
Fig.~1(a-c)), is
\bqn
\frac{1}{\sigma_0} d \sigma^{(4)} = \left( \frac{\as C_F}{\pi} \right)^2
          \left[ F_A (y_{ij}) +\left( 1 - \frac{1}{2} \frac{N_C}{C_F} \right)
          F_B(y_{ij}) + \frac{N_C}{C_F} F_C(y_{ij}) \right]
\label{qqgg}
\eqn
and for $Z\to q\bar{q}q\bar{q}$ (Fig.\ 1d)
\bqn
\frac{1}{\sigma_0} d \sigma^{(4)}=\left( \frac{\as C_F}{\pi} \right)^2
n_f \left[ \frac{T_F}{C_F} F_D(y_{ij})+
   \left( 1 - \frac{1}{2} \frac{N_C}{C_F} \right)F_E(y_{ij})\right],
\label{qqqq}
\eqn
where $y_{ij}=m_{ij}^2/s$ denotes the scaled invariant mass squared between
a pair of partons and $n_f$ is the number of active flavours.
The colour factors are determined from the SU(3) generators $(T^a)_{ij}$ and
structure constants $f^{abc}$:
\begin{eqnarray}
\sum_a \left( T^a T^{\dagger a}\right)_{ij}  & = & \delta_{ij} C_F, \\
\sum_{a,b} f^{abc} f^{abd*} & = &  \delta^{cd}N_C, \\
{\rm{Tr}}\left[ T^a T^{b\dagger} \right]  & = &  \delta^{ab} T_F.
\end{eqnarray}
The analytical form of the functions $F_A, \ldots F_E$ can be found in
Ref.~\cite{ert}.
In the ALEPH analysis \cite{aleph}, a fit  with  $\ycut=0.03$ gives
\bqna
T_F/C_F=0.58\pm 0.17_{\rm stat} \pm 0.23_{\rm syst}, \label{tf-exp} \\
N_C/C_F=2.24\pm 0.32_{\rm stat} \pm 0.24_{\rm syst},\label{nc-exp}
\eqna
which is in good agreement with the theoretical expectation, for $n_f =
5$,
\bqna
(T_F/C_F)_{QCD}&=&0.375, \label{tf-th} \\
(N_C/C_F)_{QCD}&=&2.25 . \label{nc-th}
\eqna
However, it is important to note that
the theoretical expressions in (\ref{qqgg}) and (\ref{qqqq}) above
are only valid for {\it massless}  quarks and should be corrected
for  massive fermions. For example, for  $\ycut=0.03$ and $m_Q=5\ \GeV$
 a $Q \bar Q$ pair effectively contributes 0.8 relative to  a
massless pair.
When the mass of the $b$ quark is taken into
account and  the contribution of a light gluino of 5 GeV is included,
the value of $N_C/C_F$ does not change but (\ref{tf-th}) becomes
\bqn
(T_F/C_F)_{\rm QCD\; +\; gluino}= 0.568.
\eqn
This result is surprisingly  close to the experimental value
(\ref{tf-exp})-(\ref{nc-exp}).
A calculation using massless quarks and gluinos would yield $0.375
\times 8/5 = 0.6$ for this quantity.
Obviously, the size of the experimental errors precludes any definitive
conclusion at  present.
All we can say is that the four-jet measurements are consistent with the
gluino hypothesis.

Since we have seen in Fig.~4 that the gluino contribution  to
 the total four-jet cross is quite small, it is worth investigating
whether shape variables can provide a further discrimination
\cite{farrar}.
If we order the jets
in the final state according to their energy, it is very likely that
the two least energetic jets  come from  the  splitting of the gluon
radiated off the quark pair which couple to the $Z$, Figs.~1,2. The
angular correlation between the plane of this soft jet pair and the more
energetic primary $q \bar q$ pair is different for the $q\bar q g g$
and $q \bar q q \bar q$ final states \cite{bengtsson}.
This can be quantified by using a modified Nachtmann-Reiter angle
$\thetanr$ \cite{bengtsson,nachtmann}, defined as the angle
between the vectors $({\bf p_1- p_3})$ and  $({\bf p_2- p_4})$, where the
three-momenta are ordered according to the energy of the jet.

The distribution in $\thetanr$ for {\it massive} fermion jets with
$E_j \sim m_j$ is in fact
rather different from the massless distribution. The difference is due
to the different helicity structure of the matrix element when
masses are included \cite{spinor}.
These new helicity structures   have the   opposite behaviour in $\thetanr$
 to the massless contributions, and the
net effect is that the shape of the distribution resembles more
that of the vector boson jets.
However, these effects are mainly confined  to very small $\ycut$,
and are not important for the region of experimental interest, {\it i.e.}
 $\ycut > O(0.01)$.
Fig.~5 shows the  $\cos\thetanr$ distribution for
the various contributions to the four-jet cross section.  The
two figures correspond to (a) $\ycut = 0.01$ and (b) $\ycut = 0.08$.
At the lower  $\ycut$ value, a distinctively  different behaviour is
observed for the quark/gluino  and gluon distributions.
On the other hand, the harder cut of $\ycut = 0.08$ in Fig.~5(b) distorts the
phase space so much that the shape of the distributions is virtually
indistinguishable, and the differences due to the gluino mass
are negligible.              Note that in this figure the different
contributions are normalised separately.
The crossover between the two types of behaviour shown in Fig.~5
occurs at $\ycut\sim 0.04$.
A $\ycut$ value smaller than this is therefore
needed to distinguish the fermion and vector boson contributions.
%\begin{figure}
%\centerline{
%\inclps{fig5a.ps}{530 pt}{643 pt}{25 pt}{160 pt}{550} }
%\vspace*{0.5cm}
%\centerline{
%\inclps{fig5b.ps}{530 pt}{643 pt}{25 pt}{160 pt}{550}  }
%\caption{Shape distribution in $\cos\thetanr$  of the 4-jet  cross section
%         for (a) $\ycut=0.01$ and
%        (b)  $\ycut=0.08$. The difference in the shape of the gluon
%         distribution is due to the influence of the hard cuts on the phase
%	space.}
%\end{figure}

Using angular distributions of this type, the OPAL collaboration
 has recently put
bounds on the production rate of 4-quark jet events \cite{opal}.
They find an upper limit
of 4.7\% at 68\% confidence level ({\it cl}) and
of 9.1\% at 95\% {\it cl} on the fraction of 4-jet events of
 fermion type.
The theoretical prediction of QCD is 4.7\% , and so that the inclusion of a
light
gluino would naively enhance this value to $4.7 \times 8/5=7.5 \%$. However,
at $\ycut=0.01$ (the value used by OPAL)  a  5 GeV fermion contributes only
 0.51 relative to   a massless
one. Therefore the production rate with a 5 GeV gluino included
 is only enhanced to a value of 5.25 \% (73\% {\it cl}).
This  weakens considerably the strength of the bounds coming from
this approach.
%\begin{figure}
%\centerline{
%\inclps{fig6.ps}{530 pt}{643 pt}{25 pt}{160 pt}{600}    }
%\caption{Four-jet  cross section differential in  $\cos\thetanr$.
%        The solid line corresponds to QCD (gluon+quarks), while the dashed
%	line corresponds to QCD+gluino.}
%\end{figure}

Finally, we  show in Fig.~6 the theoretical predictions for the
$\cos\thetanr$ distributions for  QCD (quark+gluons)
and  for  QCD+gluino, with $\ycut = 0.01$ and $\mg = 5\ \GeV$.
 Due to the order of magnitude difference
between the quark and gluon contributions (Fig.~4)
the differences are rather small.
Even with improved statistics,  the
procedure of Ref.~\cite{opal}  will have difficulty in putting
stringent bounds on the existence of light gluinos.

\subsection*{Acknowledgements}
We are grateful to David Summers and Siggi Bethke  for useful discussions.
R.M.T.\ acknowledges a EC post-doctoral fellowship and financial help
from CICYT contract AEN90-0033.

\newpage
\begin{description}
\item{Fig.1} Feynman diagrams contributing to the four-jet cross section.
             Other permutations are not shown.
\item{Fig.2} Feynman diagram corresponding to the the production of a pair of
             gluinos in four-jet events. The other permutation is not shown
\item{Fig.3} Mass dependence of the 4-jet total cross section for
             different $\ycut$. The values are normalized to the
             lowest order cross section.
\item{Fig.4} $\ycut$ dependence of the total cross section for
             the gluon, quark, and gluino contributions to the 4-jet final
             state.
             In the quark line, the
             mass effect of the $b$-quark is taken into account.
\item{Fig.5} Shape distribution in $\cos\thetanr$  of the 4-jet  cross section
             for (a) $\ycut=0.01$ and
             (b)  $\ycut=0.08$. The difference in the shape of the gluon
             distribution is due to the influence of the hard cuts on the phase
	     space.
\item{Fig.6} Four-jet  cross section differential in  $\cos\thetanr$.
             The solid line corresponds to QCD (gluon+quarks), while the dashed
	     line corresponds to QCD+gluino.
\end{description}
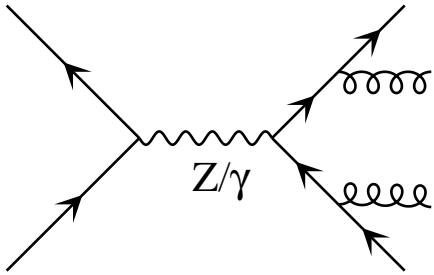
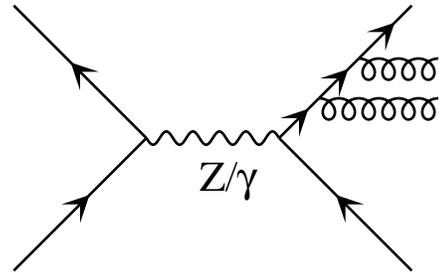
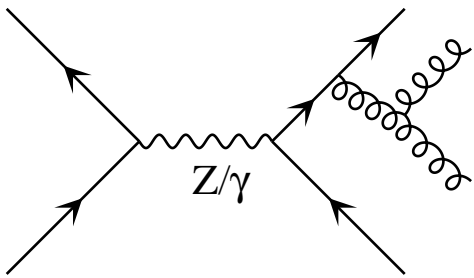
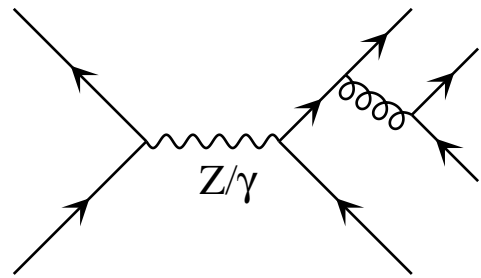
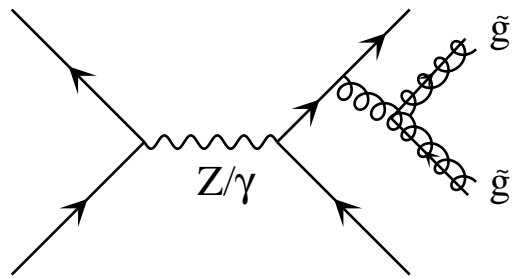
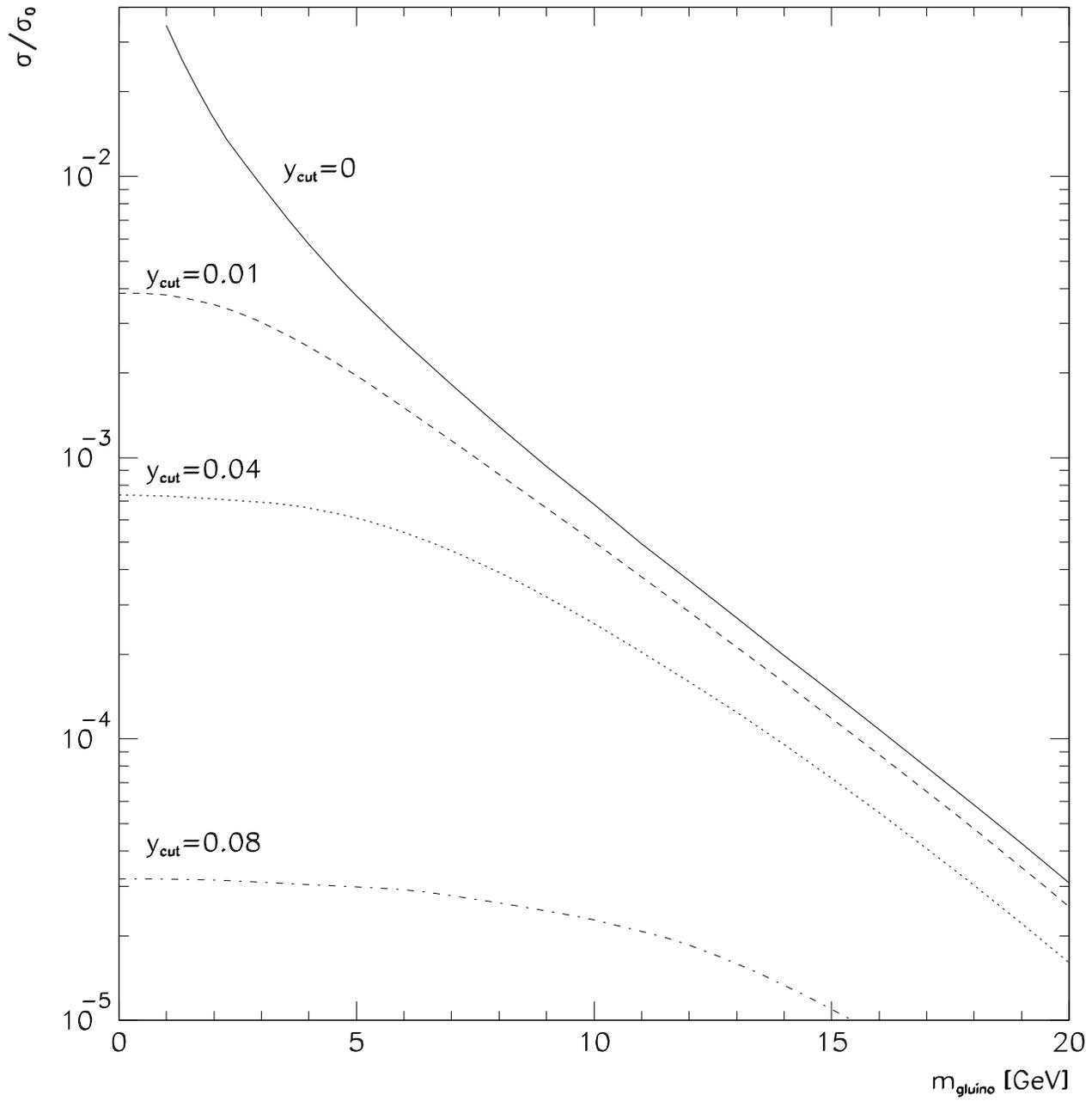
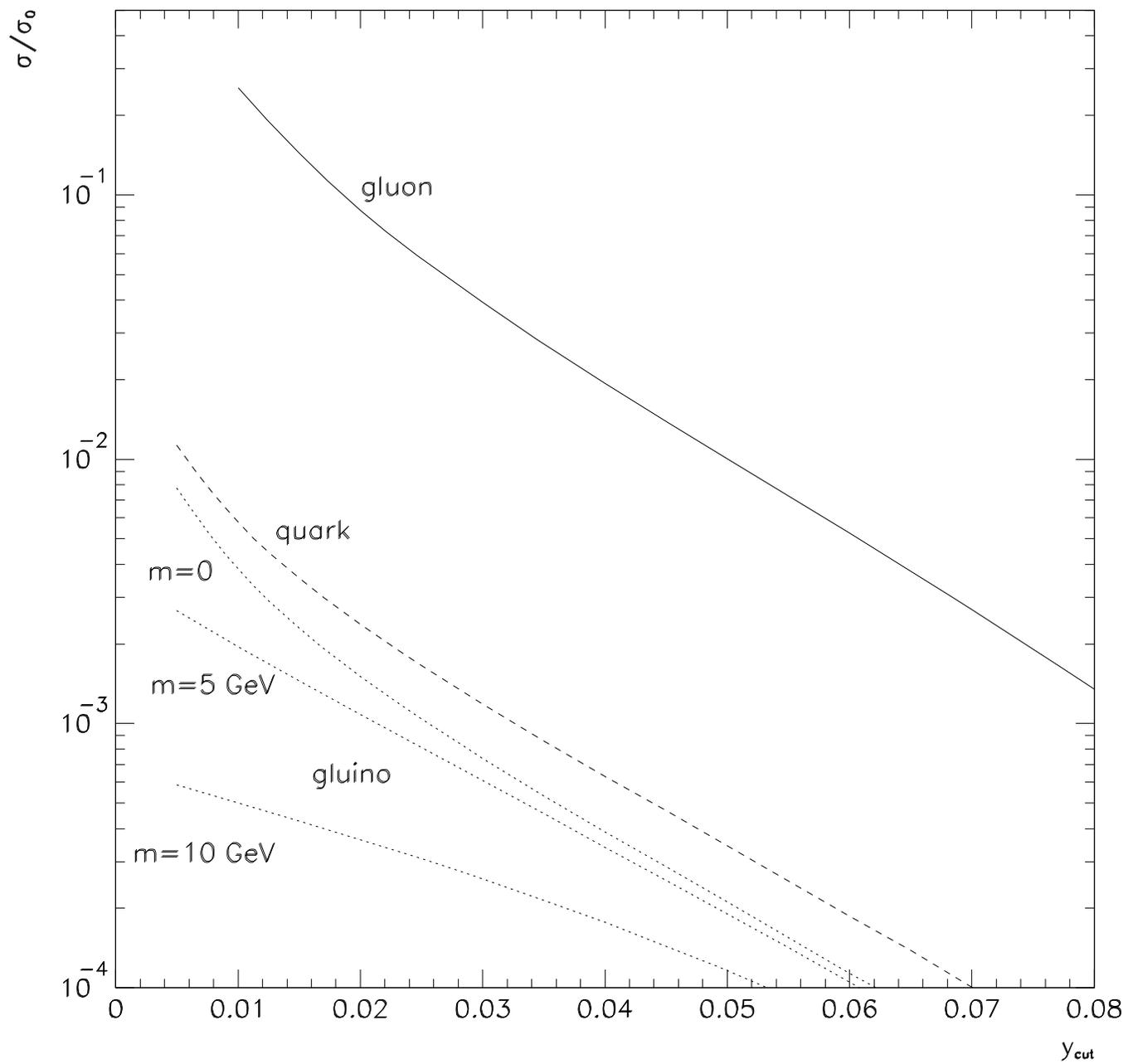
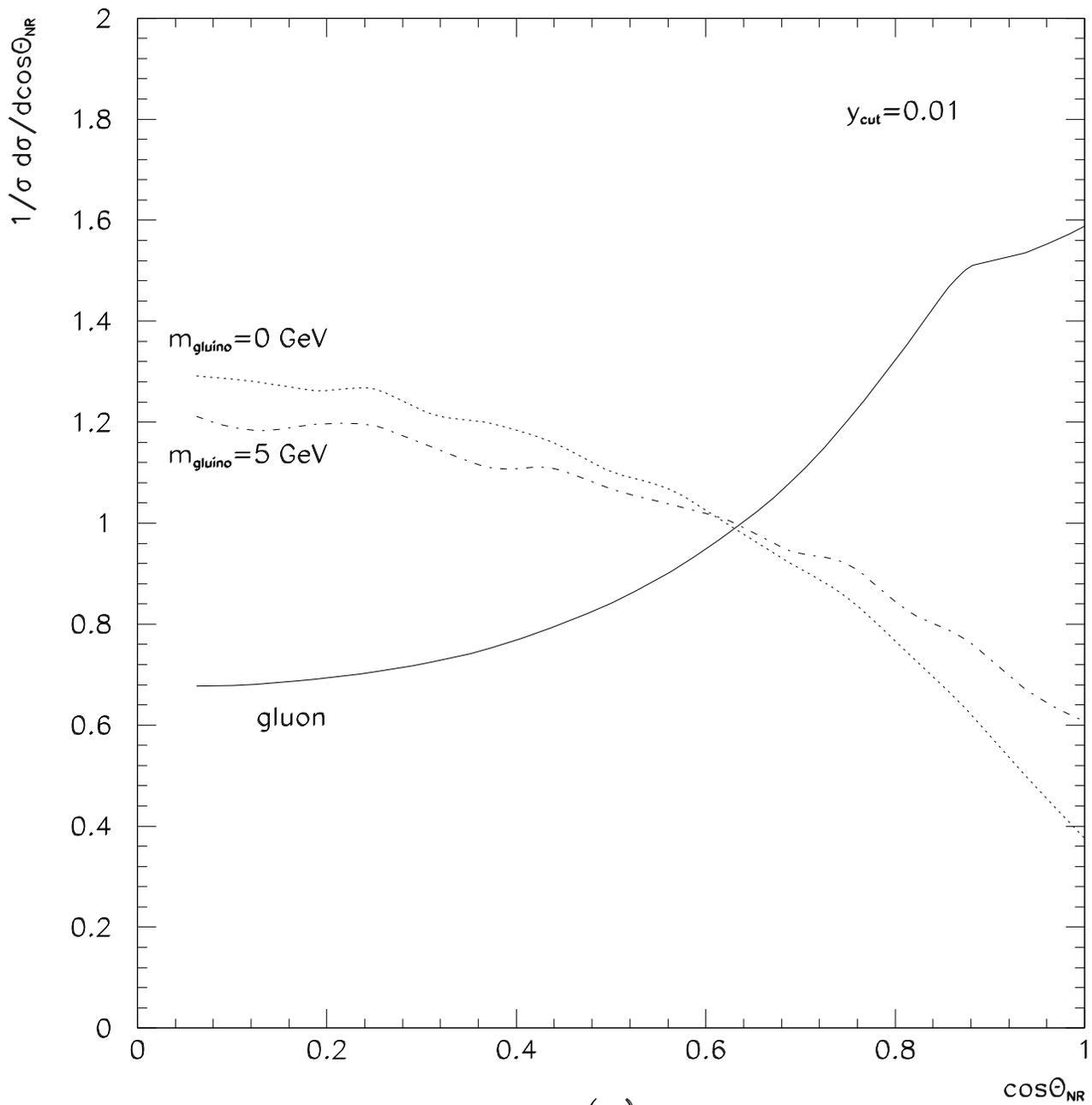
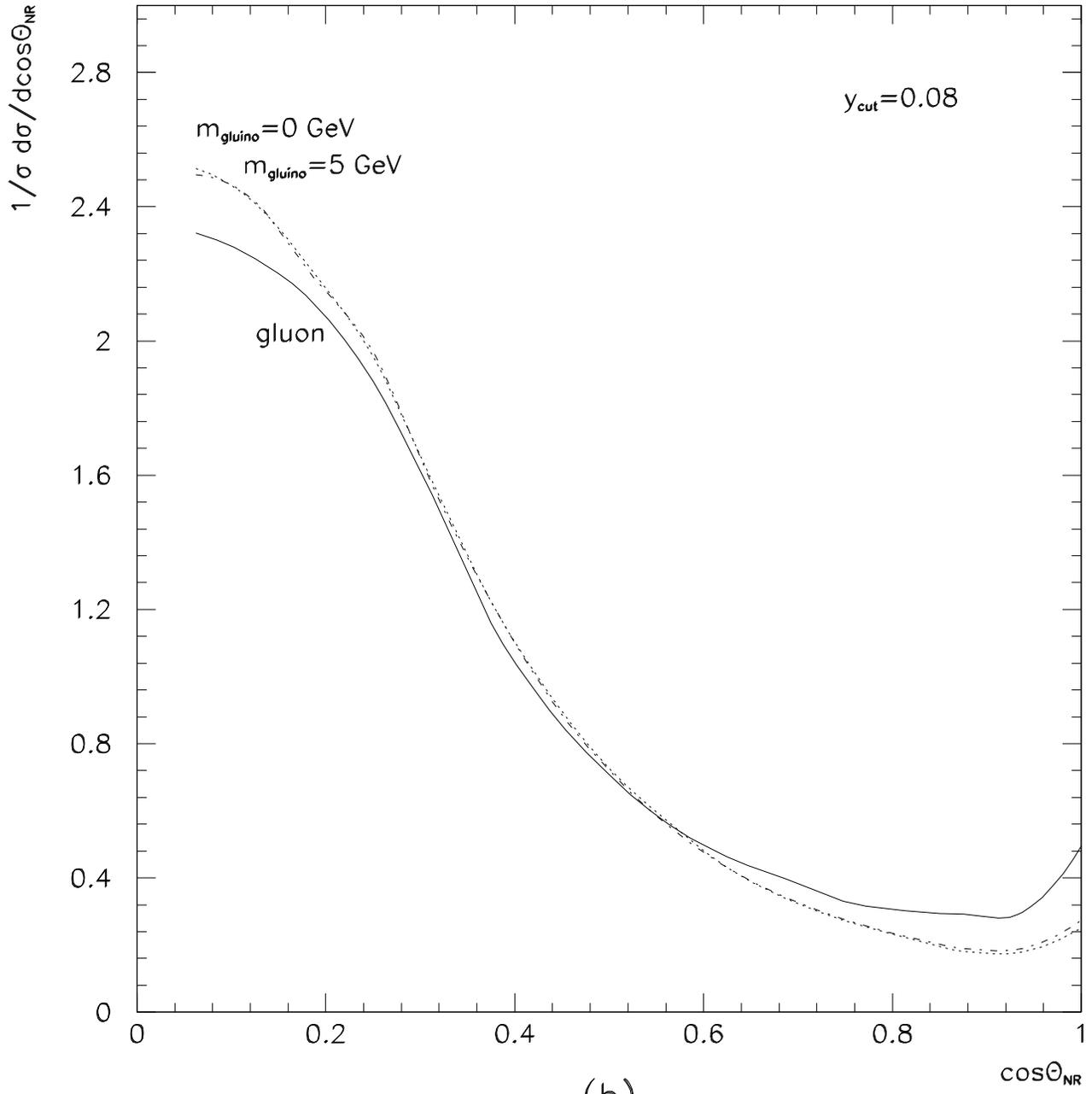
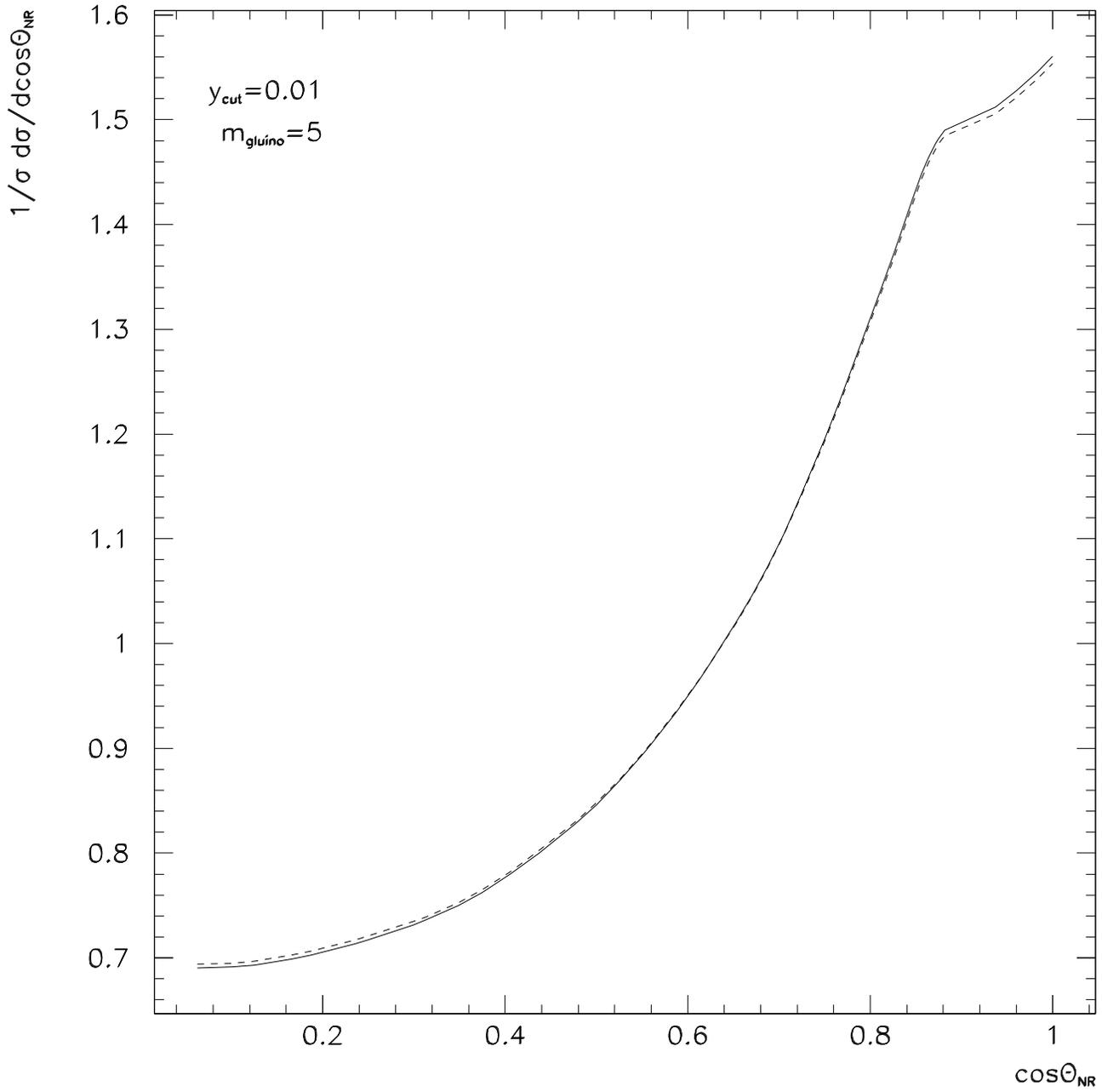
\end{document}